\documentclass[aps,prb,twocolumn]{revtex4-1}
\usepackage{amsmath,psfrag,bm}
\usepackage{xcolor}
\usepackage[caption=false]{subfig}
\usepackage{color,soul}
\usepackage{cancel}
\usepackage{epstopdf}
\usepackage{pstool}
\usepackage{etoolbox}
\usepackage{changes}
\usepackage[titletoc]{appendix}
\usepackage{float}
\begin{document}

\title{Wave Deflection and Shifted Refocusing \\ in a Medium Modulated by a Superluminal Rectangular Pulse}

\author{Zo\'e-Lise Deck-L\'eger, Alireza Akbarzadeh and Christophe Caloz }

\affiliation{Dpt. of Electrical Engineering,
Polytechnique Montr\'eal, Montr\'eal, QC H3T 1J4, Canada}
\setcounter{tocdepth}{1}

\newcommand{\ud}{\,\mathrm{d}}
\newcommand{\D}{\,\partial}
\newcommand{\rect}{\,\mathrm{rect}}
\newcommand{\sinc}{\,\mathrm{sinc}}

\begin{abstract}
We explore the problem of scattering in a medium modulated by a superluminal rectangular pulse, with the pulse modulation realized through transverse excitations. We solve this problem in the moving frame where the modulation appears purely temporal, since the solution to a purely temporal modulation is known. Upon this basis, we use a graphical approach to find the angle and frequency of the wave scattered by the modulation under oblique incidence. This study reveals that superluminal modulations deflect plane waves retro-directively, and refocus cylindrical waves to a shifted point of space with respect to the original source.
\end{abstract}

\maketitle

\section{Introduction} Specular reflection from water or smooth surfaces has been experienced by humans since the dawn of times. The reflection formula, $\theta_\text{r}=\theta_\text{i}$, was first reported in the book `Catoptrics', presumably attributed to Euclid~\cite{heath1921history}. Reflection (and refraction) from slabs and more complex structures were then studied by Huygens, Newton and many others.

\begin{figure}[!h]
\centering
\includegraphics[width=1\columnwidth]{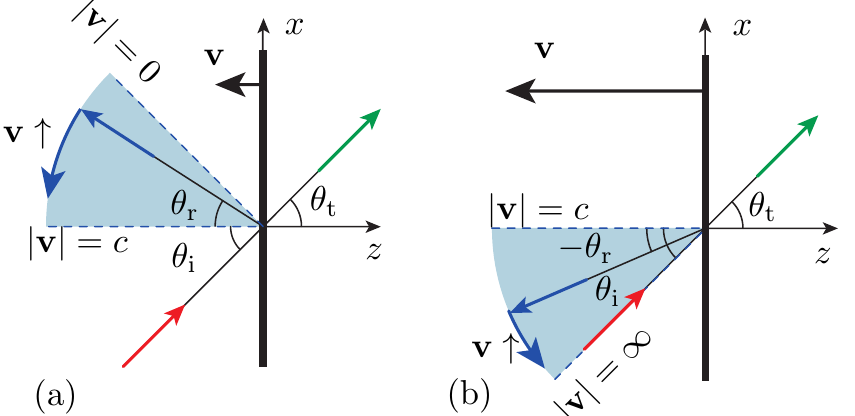}
\centering
\caption{Deflection from a rectangular pulse modulation. (a)~Subluminal (b)~Superluminal.}
\label{fig:reflection_snell}
\end{figure}

In 1727, Bradely discovered that the aberration in the perceived positions of the stars was due to the motion of the earth~\cite{bradley1727letter}, and subsequently established the theory of aberration of light. In his foundational 1905 paper on the theory of relativity~\cite{einstein1905elektrodynamik}, Einstein provided the relativistic correction to Bradley's aberration law, and solved the problem of reflection by a moving mirror, which corresponds to a double-aberration problem. Specifically, he provided formulas for the reflection angle and the reflection frequency~\cite{doppler1842}. It was later shown that the reflection phenomenology is the same for a moving dielectric slab, and the transmitted wave emerges at the same angle as the incident one~\cite{Yeh1966}. The problem of reflection by a moving slab is depicted in Fig.~\ref{fig:reflection_snell}, with the reflected wave progressively deflected from the specular angle ($\theta_\text{r}=\theta_\text{i}$) for $|\mathbf{v}|=0$ to the normal of the interface ($\theta_\text{r}=0$) for $|\mathbf{v}|=c$.

Einstein's work on the moving mirror problem was restricted to subluminal velocities, since he was considering a \emph{moving} medium, whose velocity is limited to $v\leq c$ from his own theory. Here, we will lift this restriction by considering \emph{modulated} media, where $v\geq c$ is achievable when the source of modulation is transverse to light propagation~\cite{pierce1958conservation}. Moving and modulated media are fundamentally different, but still exhibit a number of similarities. Isotropic moving media appear bianisotropic to a rest observer~\cite{kunz1980plane}, whereas modulated isotropic media remain isotropic. Moreover, moving media induce Fizeau drag~\cite{fizeau1851hypotheses}, while modulated media do not. However, both media induce Doppler frequency shifting and angle scanning.

The limiting case of a superluminally modulated medium is a purely temporally modulated medium, for which $v=\infty$. This corresponds to a medium that is switched everywhere in space between two states. There has been substantial research on such media. First, Morgenthaler~\cite{morgenthaler1958} solved the problem of scattering by a temporal step modulation and derived the corresponding frequency and amplitude scattering formulas~\footnote{He did this by applying continuity conditions on the $D$ and $B$ fields, while other authors (e.g.~\cite{kalluri2010electromagnetics}) considered continuity of the $E$ and $H$ fields. The nature of the conserved fields does not affect the results reported in this paper.}. Shortly later, Felsen noted that a temporal step modulation refocusses reflected waves back to their source position~\cite{felsen1970wave}, which was recently experimentally demonstrated with an acoustic wave in~\cite{fink2016}. Kalluri addressed the problem of temporal modulation in plasmas; he published a textbook on the topic~\cite{kalluri2010electromagnetics} and recently reported the application of a microwave to terahertz frequency and polarization transformer~\cite{kalluri2012frequency}. Finally, Halevi theoretically studied periodic temporal media~\cite{halevi2009} and experimentally demonstrated them in microwave transmission lines loaded by time-modulated varactors~\cite{halevi2016electromagnetic}.

Superluminally modulated media, although potentially harboring much richer physics than their purely temporal counterparts, due to the additional momentum parameter, have been much less explored to date. The few studies on the topic have all been restricted to waves normally incident to the superluminal modulation. Pierce and Ostrovski{\u\i}~\cite{pierce1958conservation,ostrovskii1975} pointed out the practical feasibility of such modulations and computed the related scattered frequencies for a step pulse modulation. Biancalana~\cite{biancalana2007dynamics} extended this work by additionally providing the scattered field amplitudes and solving the problem of multiple rising and falling edges (for normal wave incidence). Cassedy studied periodic superluminal modulations; specifically, he derived their dispersion relations, plotted the corresponding oblique dispersion diagrams and described their instability regions~\cite{olinercassedy1967}. Finally, new electromagnetic modes occurring in dispersive media modulated by a periodic superluminal wave were recently reported in~\cite{chamanara2017new}.

In this paper, we solve the problem of scattering from a superluminal rectangular pulse modulation. For this purpose, we first provide a graphical solution for the case of normal incidence. We particularly show that the superluminal modulation requires a time-like description, rather than the conventional space-like one for the subluminal case. We next extend the graphical approach to the case of oblique incidence. Specifically, we show that the reflected wave scans negative angles, as illustrated in Fig.~\ref{fig:reflection_snell}(b), which extends the overall scanning range beyond the normal of the modulation compared to the subluminal case [Fig.~\ref{fig:reflection_snell}(a)], and we derive formulas for corresponding scattered angles and frequencies. Finally, we demonstrate that cylindrical waves are refocused by the superluminal rectangular pulse modulation to a point that is shifted from the position of the original source. We focus on a rectangular pulse rather than a simple step pulse because this provides an opportunity to highlight a multiple scattering phenomenology that is fundamentally different from that occurring in the subluminal regime. This also allows convenient comparison with subluminal reflection, as will be seen.

\section{Graphical Analysis} 
\begin{figure}
\centering
\includegraphics[width=1\columnwidth]{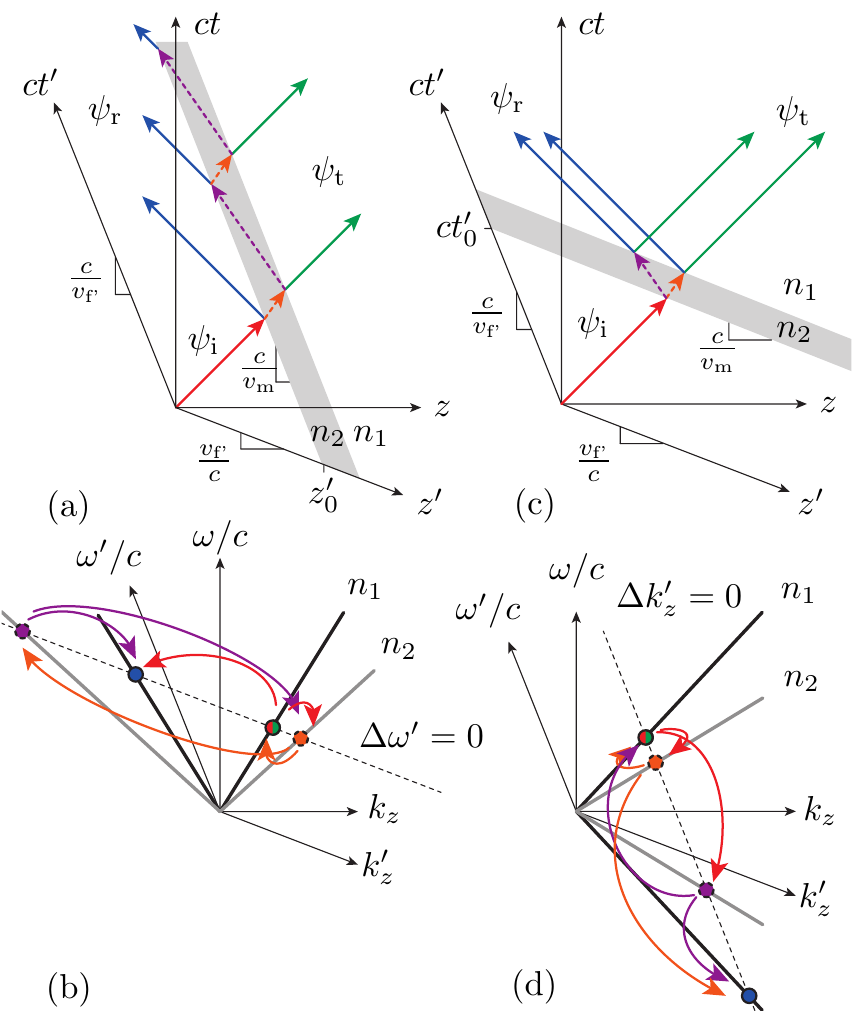}
\centering
\caption{Scattering from subluminal and superluminal rectangular pulse modulations for \emph{normal} incidence in direct and inverse Minkowski diagrams. (a)~Subluminal regime, direct space. (b)~Superluminal regime, direct space. (c)~Subluminal regime, inverse space. (d)~Superluminal regime, inverse space. The unprimed and primed axes labels correspond to the stationary and moving frames, respectively.}
\label{fig:LT}
\end{figure}

Figure~\ref{fig:LT}(a) represents the scattering of a normally incident wave, $\psi_\text{i}$, by a subluminal rectangular pulse which modulates a medium of refractive index $n_1$ to $n_2$, in the direct Minkowski diagram. Such scattering involves multiple reflections between the rising edge and the falling edge of the modulation, which result in a global reflected wave, $\psi_\text{r}$, and a global transmitted wave, $\psi_\text{t}$. Note that the slope of the rising and falling edges of the subluminal modulation is $\D(ct)/\D z = c/v_\text{m}>1$ in this representation. In order to determine the scattered frequencies, we will solve the scattering problem in the frame moving with the modulation, where the modulation appears stationnary, as done conventionally. The space and time axes of the moving frame are set accordingly, with the rising edge trajectory fixed at $z'=z_0'$, and the time axis of the moving frame $ct'$ parallel to the modulation's trajectory. The quantities in the moving frame are related to the quantities in the stationary frame through the Lorentz transformations~\cite{einstein1905elektrodynamik}
\begin{equation}\label{eq:LT}
z'=\gamma\left(z-\beta ct\right), \qquad ct'=\gamma\left(ct-\beta z\right),
\end{equation}
with the Lorentz factor $\gamma=\left(1-\beta^2\right)^{-1/2}$ and the normalized velocity of the moving frame $\beta=v_\text{f'}/c$. The slopes of the moving axes $z'$ and $ct'$ are found by setting $ct'=0$ and $z'=0$ in~\eqref{eq:LT}, respectively.

The scattered frequencies are found in the inverse Minkowki diagram, as depicted in Fig.~\ref{fig:LT}(b), which contains the dispersion relations of the initial medium and the modulated region. The inverse Lorentz transformations~\cite{kong1975theory}
\begin{equation}\label{eq:LT_inv}
k_z'=\gamma\left(k_zz-\beta \omega/c\right), \qquad \omega'/c=\gamma\left(\omega/c-\beta k_z\right),
\end{equation}
provide the moving frame axes, that turn out to be parallel to those of the direct space.

Since the modulation is stationary in the moving frame, the frequency in this frame is conserved, i.e. $\Delta \omega'=0$ (see Appendix~\ref{sec:freq_conservation_sub}), and the multiple spectral transitions~\cite{yu2009complete} starting from the incident wave (half red circle) occur along the corresponding (dotted) oblique line, parallel to $k_z'$.

The final reflected wave (blue circle) is geometrically found to be upshifted to
\begin{equation}
\omega_\text{r}=\omega_\text{i}(1-v_\text{m}^2/c^2)/(1+v_\text{m}^2/c^2),
\end{equation}
with $v_\text{m}<0$, in accordance with the Doppler law, while the transmitted wave (half green circle) is found to have the same frequency as the incident wave, $\omega_\text{t}=\omega_\text{i}$. Note that the final frequency solutions are independent of the refractive index of the modulated region.

Scattering by a superluminal rectangular pulse modulation is represented in the direct Minkowski diagram Fig.~\ref{fig:LT}(c), with the slope of the modulation trajectory being now $\D(ct)/\D z = c/v_\text{m}<1$. The incident wave, $\psi_\text{i}$, scatters into two waves at the rising edge. The reflected wave is in the second medium, contrary to the subluminal case, since it is now slower than the interface. At the falling edge, each wave generates in turn two scattered waves, for a total of four scattered waves, as in the case of the purely temporal rectangular modulation~\cite{kalluri2010electromagnetics}. This suggests that that the superluminal problem is like a temporal problem, or ``time-like'', and should consequently be transformed into a temporal problem, rather than a spatial one as done in the space-like subluminal case. \footnote{Note that the Lorentz factor ($\gamma$) becomes imaginary for $v_{\text{f}'}>c$. Therefore, transforming the superluminal problem into a spatial problem would yield unphysical imaginary space and time quantities, which we avoid by transforming to a temporal modulation.} The moving frame axes are set accordingly, with the pulse rising edge fixed at $ct'=ct_0'$, and the moving space axis parallel to its trajectory. Since the modulation is now parallel to $z'$, we find, by inspection of Fig.~\ref{fig:LT}(c), the following fundamental relationship between the modulation and frame velocities:
\begin{equation}\label{eq:v_inverse}
v_\text{f'}/c=c/v_\text{m},
\end{equation}
where $v_\text{f'}$ is the velocity an observer must have to see the modulation moving at $v_\text{m}'=\infty$, or to see the subluminal modulation as temporal.

The graphical resolution for the superluminal rectangular pulse modulation is provided in Fig.~\ref{fig:LT}(d). Since the modulation is purely temporal in the moving frame, the wavenumber in this frame is conserved, i.e. $\Delta k_z'=0$ (see Appendix~\ref{sec:freq_conservation_sup}), and the spectral transitions from the incident wave (half red circle) hence occur along the corresponding (dotted) oblique line, parallel to $\omega'/c$ this time. This indicates that reflected waves now correspond to negative frequencies, rather than negative wavenumbers. The four scattering events lead to the final reflected wave (blue circle) being upshifted to
\begin{equation}
\omega_\text{r}=-\omega_\text{i}(1-v_\text{m}^2/c^2)/(1+v_\text{m}^2/c^2),
\end{equation}
 and the transmitted wave (half green circle) having the same frequency as the incident wave, $\omega_\text{t}=\omega_\text{i}$. Once again, the final frequency solutions are independent of the refractive index of the modulated region. 

\section{Reflection from Superluminal Pulse Modulation} 
\begin{figure}[!h]
\centering
\includegraphics[width=0.7\columnwidth]{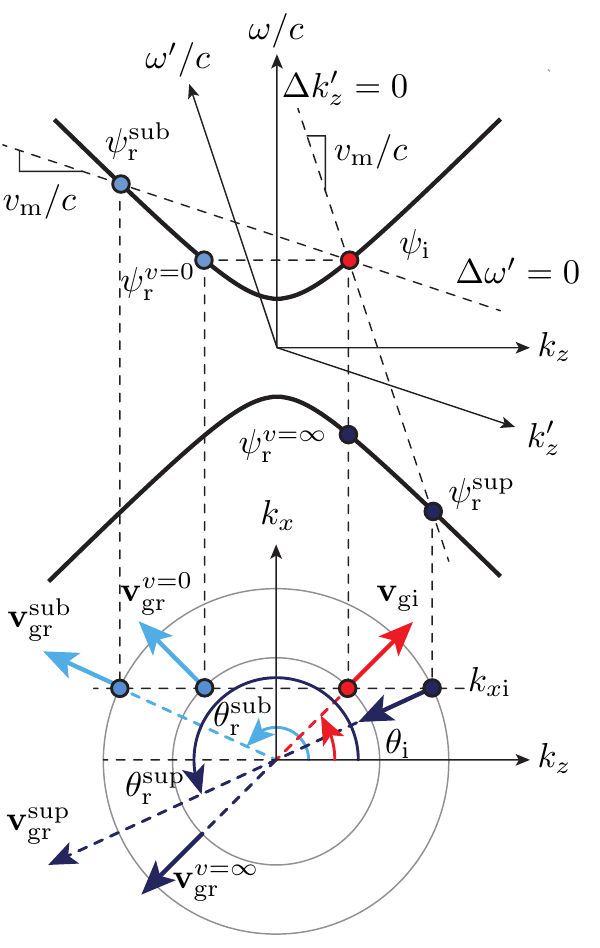}
\centering
\caption{Scattering of harmonic plane waves from subluminal and superluminal rectangular pulse modulations for \emph{oblique} incidence.  Top: iso-$k_x$ dispersion diagram. Bottom: iso-$\omega$ diagram. This graph is for the particular case of $v_\text{sub}/c=c/v_\text{sup}$, which conveniently leads to shared Lorentz moving frames. Note that we have here changed the convention angle of Fig.~\ref{fig:reflection_snell}, for convenience.}
\label{fig:deflection_disp}
\end{figure}

We now extend the problems studied in the previous section to the case of oblique incidence. Consider an incident wave propagating at an angle $\theta_\text{i}$ (Fig.~\ref{fig:reflection_snell}) with frequency $\omega_\text{i}$ and wavenumbers $k_{z\text{i}}=(\omega_\text{i}/c)\cos\theta_\text{i}$, $k_{x\text{i}}=(\omega_\text{i}/c)\sin\theta_\text{i}$. Figure~\ref{fig:deflection_disp} plots the hyperbolic dispersion curve $\omega^2/c^2-k_z^2=k_{x\text{i}}^2$ (top), with incidence point ($k_{z\text{i}},\omega_\text{i}$) (red circle), and the corresponding isofrequency curve $k_z^2+k_x^2=\omega_\text{i}^2/c^2$ (bottom), with incidence point ($k_{z\text{i}},k_{x\text{i}}$) and group velocity $\mathbf{v}_\text{g\text{i}}=\nabla_\mathbf{k}\omega_{\text{i}}(\mathbf{k})$.

We first consider the subluminal case. Due to phase matching, $k_{x\text{r}}=k_{x\text{i}}$, and therefore the reflected wave, $\psi_\text{r}^\text{sub}$, lies on the same hyperbola as the incident wave. Enforcing $\Delta\omega'=0$ (space-like problem), we localize the frequency $\omega_{\text{r}}^\text{sub}$ of the reflected wave in the $(k_z,\omega/c)$ space at the intersection of the dispersion curve and the dotted line parallel to $k_z'$ (pale blue circle). We next project this point onto the circle $k_z^2+k_x^2=(\omega_\text{r}^{\text{sub}})^2/c^2$ in the $(k_z,k_x)$ space, and find the propagation direction of the reflected wave as $\mathbf{v}_\text{g\text{r}}^{\text{sub}}=\nabla_\mathbf{k}\omega_\text{r}^{\text{sub}}(\mathbf{k})$.
We then immediately see that as $v_\text{m}$ ($\propto$ slope) increases, $\omega_\text{r}^{\text{sub}}$ increases and hence the angle of reflection, $\theta_\text{r}^{\text{sub}}$, increases towards the normal of the modulation, in agreement with Einstein's mathematical result [Fig.~\ref{fig:reflection_snell}(a)], from $\theta_\text{r}^{\text{sub}}(v_\text{m}=0)=\pi-\theta_\text{i}$ to $\theta_\text{r}^{\text{sub}}(v_\text{m}=c)=\pi$.

Now consider the superluminal case. Enforcing \mbox{$\Delta k_z'=0$} (time-like problem) yields the frequency $\omega_{\text{r}}^\text{sup}$ of the reflected wave, $\psi_\text{r}^\text{sup}$, in the $(k_z,\omega/c)$ space at the intersection of the dispersion curve and the dotted line parallel to $\omega'/c$ (dark blue circle). Projecting this point onto the $(k_z,k_x)$ space to the circle $k_z^2+k_x^2=(\omega_\text{r}^{\text{sub}})^2/c^2$ yields the propagation direction of the reflected wave $\mathbf{v}_\text{g\text{r}}^{\text{sup}}$. We then see that as $v_\text{m}$ increases, $|\omega_\text{r}^{\text{sub}}|$ \emph{decreases} and hence the angle of reflection, $\theta_\text{r}^{\text{sup}}$, is \emph{further increased} towards the incidence direction, from $\theta_\text{r}^{\text{sub}}(v_\text{m}=c)=\pi-\theta_\text{i}$ to $\theta_\text{r}^{\text{sub}}(v_\text{m}=\infty)=\pi+\theta_\text{i}$. This phenomenon, illustrated in Fig.~\ref{fig:reflection_snell}(b), constitutes a central result of the paper, and extends Einstein's law to the superluminal case.

We now mathematically derive the quantitative results corresponding to the graphical qualitative results for the superluminal modulation. In a purely temporal modulation, corresponding to the limit $v_\text{m}=\infty$ in Fig.~\ref{fig:deflection_disp}, the reflected wave is aligned with the incident wave~\cite{felsen1970wave}, i.e. $\theta_\text{r}=\theta_\text{i}+\pi$. The same is true for the superluminal modulation in the frame where it is temporal, i.e. $\theta_\text{r}'=\theta_\text{i}'+\pi$. Therefore, we have, with $v_{x\text{i,r}}'=c\sin\theta'_{\text{i,r}}$, $v_{x\text{i}}'= -v_{x\text{r}}'$. Upon relativistic transformation of velocities~\cite{einstein1905elektrodynamik} (see Appendix~\ref{sec:deflection_v}), this equation becomes in the stationary frame
\begin{equation}
  v_{x\text{r}}=\frac{v_{x\text{i}}(v_\text{f'}^2/c^2-1)}{(v_\text{f'}^2/c^2-2v_zv_\text{f'}/c^2+1)}
\end{equation}
which, with $v_{x\text{i,r}}=c\sin\theta_\text{i,r}$, translates into
\begin{equation}\label{eq:sin_sup}
  \sin\theta_\text{r}=\frac{(v_\text{f'}^2/c^2-1)\sin\theta_\text{i}}
  {v_\text{f'}^2/c^2-2\cos\theta_\text{i}v_\text{f'}/c+1}.
\end{equation}
The relations between the reflection angle and the modulation velocity are found by inserting~\eqref{eq:v_inverse} into~\eqref{eq:sin_sup}.
\begin{subequations}\label{eq:cos_sin_sup}
\begin{equation}\label{eq:sin_sup2}
   \sin\theta_\text{r}=
   \frac{(1-v_\text{m}^2/c^2)\sin\theta_\text{i}}
   {1-2\cos\theta_\text{i}v_\text{m}/c+v_\text{m}^2/c^2},
\end{equation}
\begin{equation}\label{eq:cos_sup}
   \cos\theta_\text{r}=-\frac{(1+v_\text{m}^2/c^2)\cos\theta_\text{i}-2v_\text{m}/c}
    {1-2\cos\theta_\text{i}v_\text{m}/c+v^2_\text{m}/c^2},
\end{equation}
\end{subequations}
where~\eqref{eq:cos_sup} was found by performing similar operations to $v_z$. These relations reveal that $\sin\theta_\text{r}<0$ and $\cos\theta_\text{r}<0$ for $|v_\text{m}|>c$, in agreement with our qualitative graphical analysis [Fig.~\ref{fig:deflection_disp}]. Moreover, the frequency of the reflected wave is found by inserting~\eqref{eq:sin_sup} into the conservation of momentum $k_{x\text{i}}=k_{x\text{r}}$, or $\omega_\text{i}\sin\theta_\text{i}=\omega_\text{r}\sin\theta_\text{r}$, which yields
  \begin{equation}\label{eq:fref}
    \omega_\text{r}=\omega_\text{i}
   \frac{(1-v_\text{m}^2/c^2)}
   {1-2\cos\theta_\text{i}v_\text{m}c+v_\text{m}^2c^2}.
  \end{equation}
Note that the relations~\eqref{eq:cos_sin_sup} and~\eqref{eq:fref} are identical to those found by Einstein for reflection by a subluminal mirror~\cite{einstein1905elektrodynamik} (Appendix~\ref{sec:sub_deflection_v}) and our finding therefore generalizes these laws to \emph{any} velocity in the case of a \emph{rectangular pulse}. This is not true in the case of a superluminal \emph{step pulse}, where the reflected wave propagates in the medium~2 [Fig.~\ref{fig:LT}(c)].

Note also the interesting symmetry occurring in the case of opposite velocities, $v_\text{m}^\text{sub}/c=c/v_\text{m}^\text{sup}$, and corresponding to the graph of Fig.~\ref{fig:deflection_disp}: $\theta_\text{r}^\text{sup}=2\pi+\theta_\text{r}^\text{sup}$, i.e. $\theta_\text{r}^\text{sup}=-\theta_\text{r}^\text{sub}$ in the angle convention of Fig.~\ref{fig:reflection_snell}, and $\omega_\text{r}^\text{sup}=-\omega_\text{r}^\text{sub}$.

\section{Superluminal Shifted Refocusing} 
\begin{figure}[!h]
\centering
\includegraphics[width=1\columnwidth]{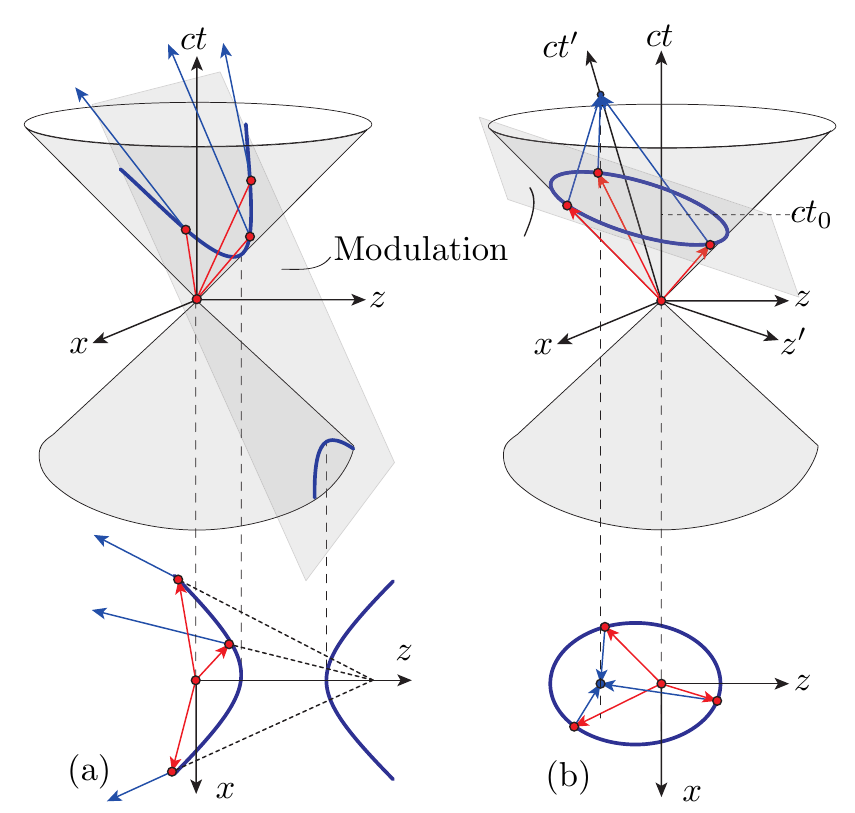}
\centering
\caption{Scattering of cylindrical wave by subluminal and superluminal pulse modulations. (a)~Subluminal case: diffraction. (b)~Superluminal case: refocusing.}
\label{fig:mink_ellipse_hyperbola}
\end{figure}

\begin{figure}
\subfloat{%
  \includegraphics[width=.3\linewidth]{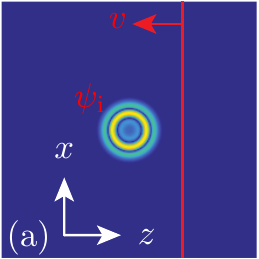}%
}\hfill
\subfloat{
  \includegraphics[width=.3\linewidth]{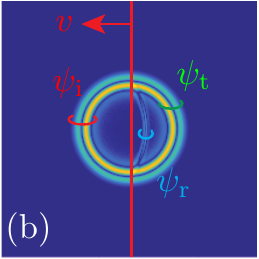}%
}\hfill
\subfloat{%
  \includegraphics[width=.3\linewidth]{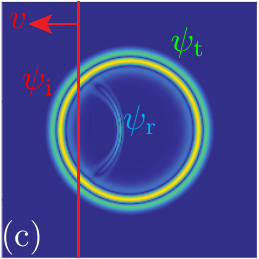}%
}\hfill
\subfloat{%
  \includegraphics[width=.3\linewidth]{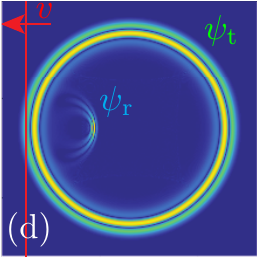}%
}\hfill
\subfloat{%
  \includegraphics[width=.3\linewidth]{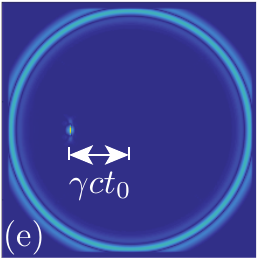}%
}\hfill
\subfloat{%
  \includegraphics[width=.3\linewidth]{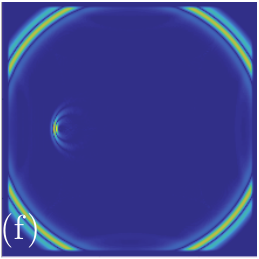}%
}\hfill
\caption{Scattering of a cylindrical pulse by a superluminal pulse modulation. (a)~Before scattering. (b)~When the modulation has reached the source point. (c)~When the modulation has crossed most of the pulse. (d)~After crossing. (e)~At the refocus time. (f)~After refocusing.\cite{supp_material}}
\label{fig:numerical}
\end{figure}

The analysis performed so far was restricted to plane waves. We now investigate the scattering of cylindrical waves, whose graphical representations are shown in Figs.~\ref{fig:mink_ellipse_hyperbola}(a) and (b) for the subluminal and superluminal cases, respectively. The intersection between the cylindrical wave and the subluminal modulation yields a hyperbola, which leads to diffraction. In contrast, the intersection between the cylindrical wave and the superluminal modulation yields an ellipse, which leads to refocusing. In the moving frame, where the modulation is seen as temporal, focusing occurs at the spatial origin of the source~\cite{felsen1970wave,fink2016}, as shown in Fig.~\ref{fig:mink_ellipse_hyperbola}(b). This is seen in the stationary frame as refocusing shifted to  the left focal point of the ellipse, the source being at the right focal point, by the amount $\gamma ct_0$.
Figure~\ref{fig:numerical} shows snapshots of a standard finite-difference time-domain (FDTD) simulation of a cylindrical pulse scattered by a narrow superluminal pulse modulation, with the modulation simulated as a change in the refractive index, updated at each time step. The incident cylindrical pulse is generated from an infinite line source perpendicular to the 2D scattering problem. Refocusing is seen to occur to the left of the origin of the source with the expected shift.

\section{Conclusion}
We have solved the canonical problem of the scattering of waves in a medium modulated by a superluminal rectangular pulse. Such a medium could be experimentally  realized in the form of a 2D planar transmission line structure~\cite{eleftheriades2002planar,sanada2004planar}, which may be modulated using fast-switching elements such as varactors~\cite{taravati2017}. Our findings may lead to a new class of devices controlling both the spatial and temporal spectra of waves.

\appendix

\section{Spatial and Temporal Frequency Conservation}\label{sec:freq_conserved}

Here, we derive the conserved frequency quantities at subluminal and superluminal modulations.

\subsection{Subluminal Case}\label{sec:freq_conservation_sub}

We start by studying the scattering from the rising edge of the rectangular pulse modulation. The incident and reflected waves are in the first (unmodulated) medium and the transmitted wave is in the second, (modulated) medium, as in Fig.~\ref{fig:LT}(a),(b), for the red, blue and dashed orange waves. The wave is a monochromatic plane wave with a $y$-polarized electric field, propagating in the $x-z$ direction, i.e.:
\begin{equation}\label{eq:E_field}
\textbf{E}_\text{m}=A_\text{m} e^{i(k_{x\text{m}}x\pm k_{z\text{m}}z-\omega_{x\text{m}})}\hat{\textbf{y}},
\end{equation}
with $\text{m}=$i, r and 2t, where 2t stands for transmitted in medium 2. In the moving frame,
\begin{equation}\label{eq:E_field_mf}
\textbf{E}'_\text{m}=A'_\text{m} e^{i(k'_{x\text{m}}x'\pm k'_{z\text{m}}z'-\omega'_{x\text{m}}t)}\hat{\textbf{y}'},
\end{equation}
The tangential electric field is continuous across the boundary in the moving frame, i.e:
\begin{equation}\label{eq:E_field_continuity}
\left[E'_{y\text{i}}+E'_{y\text{r}}=E'_{y\text{2t}}\right]_{z'=0}.
\end{equation}
Inserting \eqref{eq:E_field_mf} into \eqref{eq:E_field_continuity} yields
\begin{equation}
\begin{split}
A'_\text{i}e^{i(k'_{x\text{i}}x'+k'_{z\text{i}}z'-\omega'_\text{i}t')}&+
A'_\text{r}e^{i(k'_{x\text{r}}x'-k'_{z\text{r}}z'-\omega'_\text{r}t')}\left.\right|_{z=0}\\
=A'_\text{2t}&e^{i(k'_{x\text{2t}}x'+k'_{z\text{2t}}z'-\omega'_\text{2t'} t)}\left.\right|_{z=0},
\end{split}
\end{equation}
from which we deduce the conservation of $k'_x$ and $\omega'$:
\begin{equation}
k'_{x\text{i}}=k'_{x\text{r}}=k'_{x\text{2t}}, \qquad \omega'_\text{i}=\omega'_\text{r}=\omega'_\text{2t}.
\end{equation}
These equations were found for the rising edge of the rectangular pulse modulation, but are also valid for falling edge, such that:
\begin{equation}
k'_{x\text{2t}}=k'_{x\text{2r}}=k'_{x\text{t}}, \qquad \omega'_\text{2t}=\omega'_\text{2r}=\omega'_\text{t}.
\end{equation}
therefore the spatial frequencies before and after the rectangular modulation are related by
\begin{equation}
k'_{x\text{i}}=k'_{x\text{r}}=k'_{x\text{t}}, \qquad \omega'_\text{i}=\omega'_\text{r}=\omega'_\text{t}.
\end{equation}
\subsection{Superluminal Case}\label{sec:freq_conservation_sup}

We now study a temporal modulation, which is the limiting case of a superluminal modulation. We start by studying the change from $n_1$ to $n_2$. The fields before and after the modulation are continuous. There is a controversy on the nature of the conserved field, as to whether the $\mathbf{D}$ or $\mathbf{E}$ field is conserved, but this does not affect our result, since we are only investigating the phase quantities and not the amplitudes. We choose to write the conservation of $\mathbf{D}$, following \cite{morgenthaler1958}, i.e.
\begin{equation}\label{eq:D_field_continuity}
\left[\mathbf{D}_\text{i}=\mathbf{D}_\text{2r}+\mathbf{D}_\text{2t}\right]_{t=0}
\end{equation}
where 2r, 2t and t correspond to the red, dashed orange and dashed purple in Fig.~\ref{fig:LT}(c),(d). Following the same procedure as above, we find
\begin{equation}
\begin{split}
A_\text{i}e^{i(k_{x\text{i}}x+k_{z\text{i}}z-\omega_\text{i}t)}\left.\right|_{t=0}&\\
=A_\text{2r}e^{i(k_{x\text{2r}}x+k_{z\text{2r}}z+\omega_\text{2r}t)}&+
A_\text{2t}e^{i(k_{x\text{2t}}x+k_{z\text{2t}}z-\omega_\text{2t} t)}\left.\right|_{t=0},
\end{split}
\end{equation}
from which we find
\begin{equation}\label{eq:k_conservation_temp}
k_{x\text{i}}=k_{x\text{2r}}=k_{x\text{2t}}, \qquad k_{z\text{i}}=k_{z\text{2r}}=k_{z\text{2t}}.
\end{equation}
A superluminal rising edge can be seen as a temporal modulation from $n_1$ to $n_2$ in a frame moving at the velocity $v_{\text{f}'}/c=c/v_\text{m}=$~\eqref{eq:v_inverse}. In this moving frame,
\begin{equation}\label{eq:Dp_field_continuity}
\left[\mathbf{D}'_\text{i}=\mathbf{D}'_\text{2r}+\mathbf{D}'_\text{2t}\right]_{t'=0}
\end{equation}
such that
\begin{equation}
k'_{x\text{i}}=k'_{x\text{2r}}=k'_{x\text{2t}}, \qquad k'_{z\text{i}}=k'_{z\text{2r}}=k'_{z\text{2t}}.
\end{equation}
and therefore, the frequencies before and after the modulation are related by:
\begin{equation}\label{eq:k_conservation_sup}
k'_{x\text{i}}=k'_{x\text{r}}=k'_{x\text{t}}, \qquad k'_{z\text{i}}=k'_{z\text{r}}=k'_{z\text{t}}.
\end{equation}

\section{Derivation of the Deflection Angle from Velocity Addition}\label{sec:deflection_v}

Here, we solve the deflection angle of a wave scattered by a subluminal and a superluminal modulation. We start with the subluminal modulation, for which the solution was provided by Einstein in~\cite{einstein1905elektrodynamik} (recalling that the frequency of a wave reflected from a moving wall is the same as that from a propagating modulation).
\subsection{Subluminal Case}\label{sec:sub_deflection_v}
Consider a wave obliquely incident on a subluminal rectangular pulse modulation. In the frame moving with the modulation, the modulation appears stationary, such that the transverse velocities are conserved and the normal velocities are inversed (see Fig.~\ref{fig:deflection_disp}):
\begin{equation}\label{eq:v_conservation}
   v_{z\text{i}}'= -v_{z\text{r}}', \qquad v_{x\text{i}}'= v_{x\text{r}}'.
\end{equation}
The relativistic velocity addition formulas for a frame moving along $z$ are~\cite{einstein1905elektrodynamik}
\begin{subequations}\label{eq:relativistic_velocity}
\begin{equation}\label{eq:vp}
    v_z'=\frac{v_z-v_\text{f'}}
    {1-\frac{v_zv_\text{f'}}{c^2}}, \qquad
    v_x'=\frac{v_x}  {\gamma\left(1-\frac{v_zv_\text{f'}}{c^2}\right)},
\end{equation}
\begin{equation}\label{eq:v}
    v_z=\frac{v'_z+v_\text{f'}}
    {1+\frac{v'_z v_\text{f'}}{c^2}}, \qquad
    v_x=\frac{v'_x}
    {\gamma\left(1+\frac{v_z v_\text{f'}}{c^2}\right)}.
\end{equation}
\end{subequations}
Writing \eqref{eq:v_conservation} in the stationary frame by inserting \eqref{eq:v} into \eqref{eq:v_conservation}, we obtain:
\begin{subequations}
\begin{align}
 v_{z\text{r}} &=\frac{v'_{z\text{r}}+v_\text{f'}}{1+\frac{v'_{z\text{r}} v_\text{f}'}{c^2}}\\
   &=\frac{-v'_{z\text{i}}+v_\text{f'}}{1-\frac{v'_{z\text{i}} v_\text{f'}}{c^2}}\\
   &=\frac{-\frac{v_z-v_\text{f'}}
    {1-\frac{v_zv_\text{f'}}{c^2}}+v_\text{f'}}
    {1-\frac{v_\text{f'}}{c^2}\frac{v_z-v_\text{f'}}
    {1-\frac{v_zv_\text{f'}}{c^2}}}\\
   &=\frac{-v_z+v_\text{f'}+v_\text{f'}(1-\frac{v_zv_\text{f'}}{c^2})}
    {(1-\frac{v_zv_\text{f'}}{c^2})-\frac{v_\text{f'}}{c^2}(v_z-v_\text{f'})}\\
   &=-\frac{v_z(1+\frac{v_{\text{f'}}^2}{c^2})-2v_\text{f'}}
    {1-2\frac{v_zv_\text{f'}}{c^2}+\frac{{v_\text{f'}}^2}{c^2}}.
\end{align}
\end{subequations}
Similarly, for $v_x$:
\begin{subequations}
\begin{align}
 v_{x\text{r}} &=\frac{v'_{x\text{r}}}
    {\gamma\left(1+\frac{v'_{z\text{r}} v_\text{'f}}{c^2}\right)}\\
   &=\frac{v'_{x\text{i}}}
    {\gamma\left(1-\frac{v'_{z\text{i}} v_\text{f'}}{c^2}\right)}\\
   &=\frac{v_x}
    {\gamma^2\left(1-\frac{v_\text{f'}}{c^2}\frac{v_z-v_\text{f'}}
    {1-\frac{v_zv_\text{f'}}{c^2}}\right)\left(1-\frac{v_zv_\text{f'}}{c^2}\right)}\\
   &=\frac{v_x}
    {\gamma^2\left(\left(1-\frac{v_zv_\text{f'}}{c^2}\right)-\frac{v_\text{f'}}{c^2}(v_z-v_\text{f'})
    \right)}\\
      &=\frac{v_x(1-v_\text{f'}^2/c^2)}
    {1-2\frac{v_zv_\text{f'}}{c^2}+\frac{v_\text{f'}^2}{c^2}}.
\end{align}
\end{subequations}
Substituting $v_x=c\sin\theta$, $v_z=c\cos\theta$, we find
\begin{subequations}\label{eq:cos_sin_sub}
\begin{equation}
   \cos\theta_\text{r}=-\frac{\cos\theta_\text{i}(1+\frac{v_\text{f'}^2}{c^2})-2v_\text{f'}/c}
    {1-2c\cos\theta_\text{i}\frac{v_\text{f'}}{c^2}+\frac{v^2_\text{f'}}{c^2}},
\end{equation}
\begin{equation}
   \sin\theta_\text{r}=\frac{ \sin\theta_\text{i}(1-v_\text{f'}^2/c^2)}
    {1-2 c\cos\theta_\text{i}\frac{v_\text{f'}}{c^2}+\frac{v_\text{f'}^2}{c^2}}.
\end{equation}
\end{subequations}

\subsection{Superluminal Case}\label{sec:sup_deflection_v}

Consider now a wave obliquely incident on a superluminal modulation. There is a moving frame in which this modulation appears at once, i.e. appears as a temporal modulation. The reflected wave is collinear with the incident wave, such that (see Fig.~\ref{fig:deflection_disp}):
\begin{equation}
   v_{z\text{i}}'= -v_{z\text{r}}', \qquad v_{x\text{i}}'= -v_{x\text{r}}'
\end{equation}
These equations are transformed to the rest frame using \eqref{eq:relativistic_velocity}:
\begin{subequations}
\begin{align}
 v_{z\text{r}} &=\frac{v'_{z\text{r}}+v_\text{f'}}{1+\frac{v'_{z\text{r}} v_\text{f'}}{c^2}}\\
   &=\frac{-v'_{z\text{i}}+v_\text{f'}}{1-\frac{v'_{z\text{i}} v_\text{f'}}{c^2}}\\
   &=\frac{-\frac{v_z-v_\text{f'}}
    {1-\frac{v_zv_\text{f'}}{c^2}}+v_\text{f'}}
    {1-\frac{v_\text{f'}}{c^2}\frac{v_z-v_\text{f'}}
    {1-\frac{v_zv_\text{f'}}{c^2}}}\\
   &=\frac{-v_z+v_\text{f'}+v_\text{f'}(1-\frac{v_zv_\text{f'}}{c^2})}
    {(1-\frac{v_zv_\text{f'}}{c^2})-\frac{v_\text{f'}}{c^2}(v_z-v_\text{f'})}\\
   &=-\frac{v_z(1+\frac{v_\text{f'}^2}{c^2})-2v_\text{f'}}
    {1-2\frac{v_zv_\text{f'}}{c^2}+\frac{v^2_\text{f'}}{c^2}}.
\end{align}
\end{subequations}
Similarly, for $v_x$:
\begin{subequations}
  \begin{align}
 v_{x\text{r}} &=\frac{v'_{x\text{r}}}
    {\gamma\left(1+\frac{v'_{z\text{r}} v_\text{f'}}{c^2}\right)}\\
   &=\frac{-v'_{x\text{i}}}
    {\gamma\left(1-\frac{v'_{z\text{i}} v_\text{f'}}{c^2}\right)}\\
   &=\frac{-v_x}
    {\gamma^2\left(1-\frac{v_\text{f'}}{c^2}\frac{v_z-v_\text{f'}}
    {1-\frac{v_zv_\text{f'}}{c^2}}\right)\left(1-\frac{v_zv_\text{f'}}{c^2}\right)}\\
   &=\frac{-v_x}
    {\gamma^2\left(\left(1-\frac{v_zv_\text{f'}}{c^2}\right)-\frac{v_\text{f'}}{c^2}(v_z-v_\text{f'})
    \right)}\\
      &=\frac{-v_x(1-v_\text{f'}^2/c^2)}
    {1-2\frac{v_zv_\text{f'}}{c^2}+\frac{v_\text{f'}^2}{c^2}}.
\end{align}
\end{subequations}
Substituting $v_x=c\sin\theta$, $v_z=c\cos\theta$, we find
\begin{equation}
   \cos\theta_\text{r}=-\frac{\cos\theta_\text{i}(1+\frac{v_\text{f'}^2}{c^2})-2v_\text{f'}/c}
    {1-2c\cos\theta_\text{i}\frac{v_\text{f'}}{c^2}+\frac{v^2_\text{f'}}{c^2}},
\end{equation}
\begin{equation}
   \sin\theta_\text{r}=\frac{ -\sin\theta_\text{i}(1-v_\text{f'}^2/c^2)}
    {1-2 c\cos\theta_\text{i}\frac{v_\text{f'}}{c^2}+\frac{v_\text{f'}^2}{c^2}}.
\end{equation}
We finally substitute $v_\text{f}'=c^2v_\text{m}$, from~\eqref{eq:v_inverse}, leading to
\begin{subequations}\label{eq:cos_sin_sup_2}
\begin{equation}
   \cos\theta_\text{r}=-\frac{\cos\theta_\text{i}(1+\frac{v_\text{f'}^2}{c^2})-2v_\text{f'}/c}
    {1-2c\cos\theta_\text{i}\frac{v_\text{f'}}{c^2}+\frac{v^2_\text{f'}}{c^2}}
\end{equation}
\begin{equation}
   \sin\theta_\text{r}=\frac{ \sin\theta_\text{i}(1-v_\text{f'}^2/c^2)}
    {1-2 c\cos\theta_\text{i}\frac{v_\text{f'}}{c^2}+\frac{v_\text{f'}^2}{c^2}}
\end{equation}
\end{subequations}

Notice that that the expressions \eqref{eq:cos_sin_sup_2} and \eqref{eq:cos_sin_sub} are identical.

\section{Derivation of the Deflection Angle from Continuity Conditions}\label{sec:deflection_k}

We now derive \eqref{eq:cos_sin_sup} using an alternative method, closely following~\cite{kunz1980plane}, who treated the subluminal case only. We start with the $z$-directed wavevector conservation in the moving frame~\eqref{eq:k_conservation_sup}.
\begin{equation}\label{eq:cont_kz}
k'_{z\text{i}}=k'_{z\text{r}}.
\end{equation}
Applying the Lorentz transformation~\eqref{eq:LT_inv} to~\eqref{eq:cont_kz} yields
\begin{equation}\label{eq:cont_kz_b}
k_{z\text{i}}-\beta \omega_\text{i}=k_{z\text{r}}+\beta \omega_\text{r}.
\end{equation}
We consider the initial and final media are free space, and so we substitue $\omega_\text{i,r}=k_\text{i,r}/c$ into \eqref{eq:cont_kz_b}.  Equations~\eqref{eq:cont_kz_b} can be rearranged as
\begin{equation}\label{eq:cont_kz_c}
k_{z\text{i}}-k_{z\text{r}}=\beta\left(k_\text{i}+k_\text{r}\right).
\end{equation}
The $x$-directed wavevector in the moving frame is
\begin{equation}\label{eq:cont_kx}
k_{x\text{i}}=k_{x\text{r}}.
\end{equation}
Upon squaring~\eqref{eq:cont_kx} and using the Helmholtz relation, we find
\begin{equation}
{k_\text{i}}^2-{k_{z\text{i}}}^2={k_\text{r}}^2-{k_{z\text{r}}}^2,
\end{equation}
or, after rearranging,
\begin{equation}\label{eq:6}
{k_{z\text{i}}}^2-{k_{z\text{r}}}^2={k_\text{i}}^2-{k_\text{r}}^2.
\end{equation}
Equation~\eqref{eq:6} is rewritten as
\begin{equation}\label{eq:cont_kx_d}
(k_{z\text{i}}-k_{z\text{r}})(k_{z\text{i}}+k_{z\text{r}})=(k_\text{i}-k_\text{r})(k_\text{i}+k_\text{r}),
\end{equation}
and dividing \eqref{eq:cont_kx_d} by \eqref{eq:cont_kz_c} yields
\begin{equation}\label{eq:div_kx_kz}
k_{z\text{i}}+k_{z\text{r}}=\frac{1}{\beta}\left(k_\text{i}-k_\text{r}\right),
\end{equation}
or
\begin{equation}\label{eq:div_2_kx_kz}
k_\text{i} -\beta k_{z\text{i}}= k_\text{r}+\beta k_{z\text{r}}.
\end{equation}
Substituting $k_{zm}=k_m\cos\theta_m$ into \eqref{eq:div_2_kx_kz} yields
\begin{equation}\label{eq:result_substitution a}
k_\text{i}(1-\beta\cos\theta_\text{i})=k_\text{r}\left(1+\beta\cos\theta_\text{r}\right),
\end{equation}
while performing the same substitution into~\eqref{eq:cont_kz_b} yields
\begin{equation}\label{eq:result_substitution b}
k_\text{i}\left(\cos\theta_\text{i}-\beta\right)=k_\text{r}(\cos\theta_\text{r}+\beta).
\end{equation}
Finally, dividing~\eqref{eq:result_substitution a} by~\eqref{eq:result_substitution b} yields
\begin{equation}
\frac{1-\beta\cos\theta_\text{i}}{\cos\theta_\text{i}-\beta}
=\frac{1+\beta\cos\theta_\text{r}}{\cos\theta_\text{r}+\beta},
\end{equation}
which rearranges to
\begin{equation}
\cos\theta_\text{r}=\frac{\cos\theta_\text{i}\left(1+\beta^2\right)-2\beta}
{1+\beta^2-2\beta\cos\theta_\text{i}}.
\end{equation}

\bibliography{ST_Super_Disc}

\end{document}